\input{aipcheck}

\documentclass[
    ,final            
  ]
  {aipproc}

\layoutstyle{6x9}

\begin{document}

\title{Double Helicity Asymmetry of Inclusive $\pi^0$ Production
in Polarized $pp$ Collisions at $\sqrt{s}=62.4$~GeV}


\author{K.~Aoki for the PHENIX Collaboration}{
  address={Department of Physics, Kyoto University, Kyoto, Kyoto, Japan, 606-8502}
}



\begin{abstract}
 The proton spin structure is not understood yet and
 there has remained large uncertainty on $\Delta g$,
 the gluon spin contribution to the proton.
 Double helicity asymmetry ($A_{LL}$) of $\pi^0$
 production in polarized $pp$ collisions is used to constrain
 $\Delta g$. In this report, preliminary results of
 $A_{LL}$ of $\pi^0$ in $pp$ collisions at $\sqrt{s}=62.4$~GeV
 measured by PHENIX experiment in 2006 is presented. It can probe
 higer $x$ region than the previously reported $\pi^0 A_{LL}$ at 
 $\sqrt{s}=200$~GeV thanks to the lower center of mass energy.
\end{abstract}

\maketitle


\section{Introduction}
The so-called ``proton spin crisis,'' initiated by the results from
the polarized deep inelastic scattering experiments,
has triggered wide effort towards the understanding of proton spin.
Despite the wide efforts, there has remained large uncertainty on $\Delta g$,
the gluon spin contribution to the proton. RHIC,
the world's first polarized proton-proton collider,
provides us an opportunity to directly probe gluons in the
proton.
Double helicity asymmetry ($A_{LL}$) of inclusive $\pi^0$ production in
polarized $pp$ collisions is sensitive to $\Delta g$ because
$\pi^0$ production is dominated by
gluon-gluon and quark-gluon interactions in the measured $p_T$ range.
PHENIX has previously reported $\pi^0 A_{LL}$
in $pp$ collisions at $\sqrt{s}=200$~GeV 
\ \cite{Boyle:2006ab} which is based on the data taken in 2005 (Run5)
and it indicates that $\Delta g$ is not large.\cite{Hirai:2006sr}
But a large uncertainty remains for large Bjorken $x\ (>0.1)$ and more statistics
are needed.
During the run in 2006 (Run6), one-week 
data taking was performed at $\sqrt{s}=62.4$~GeV.
Spin rotator commissioning was successful and we had longitudinally
polarized collisions. \cite{Togawa:1}
Even in this short data taking with a small integrated
luminosity of 60~nb$^{-1}$ and the average polarization of 48\%,
the data has a big advantage to cover the larger $x$ region thanks
to the lower center of mass energy.
According to a pertubative
QCD (pQCD) calculation, collisions at $\sqrt{s}=62.4$~GeV has $\sim 300$ times
larger cross-section than that at $\sqrt{s}=200$~GeV at fixed $x_T = 2p_T / \sqrt{s}$.
It corresponds to 10 times larger statistics than the previously
reported $\pi^0 A_{LL}$ which is based on the integrated luminosity of 1.8 
pb$^{-1}$ with average polarization of 47\%.

$A_{LL}$ is defined as
\begin{equation}
A_{LL} = \frac{\sigma_{++}-\sigma_{+-}}{\sigma_{++}+\sigma_{+-}}
\end{equation}
where $\sigma_{++(+-)}$ is the production cross-section in like (unlike)
helicity collisions. Experimentally, $A_{LL}$ is calculated as
\begin{equation}
 A_{LL} = \frac{1}{|P_B||P_Y|}
 \frac{N_{++}-RN_{+-}}{N_{++}+RN_{+-}},\ \
 R=\frac{L_{++}}{L_{+-}}
 \label{eq:all}
\end{equation}
where $P_{B(Y)}$ denotes the beam polarization, $N^{++(+-)}$ is the
$\pi^0$ yield and $L^{++(+-)}$ is the luminosity
in like (unlike) helicity collisions. $R$ is the relative luminosity.

\section{Experiment}
The stable polarization direction of RHIC beam is transverse.
Then it is rotated to get longitudinally polarized collisions
just before the PHENIX interaction point.
PHENIX local polarimeter\cite{Togawa:1} confirms that
the beam is longitudinal by measuring $A_N$ of forward neutrons.

PHENIX has Beam-Beam Counter (BBC) which covers $3.0<|\eta|<3.9$ and
Zero Degree Calorimeter (ZDC) which covers very forward angle
($\pm 2$mrad).\cite{Adcox:2003zm}
These two detectors serve as independent luminosity measure.
We used BBC counts to measure relative luminosity $R$ in equation
(\ref{eq:all}) and its uncertainty is estimated by
comparing to ZDC counts. It is found to be $\delta R=1.3 \times 10^{-3}$.
This corresponds
to $\delta A_{LL}=2.8 \times 10^{-3}$ which is less than
the statistical uncertainty.

PHENIX has
the ability to clearly identify $\pi^0$ through its gamma decay
by using an Electro-Magnetic Calorimeter (EMCal) which covers
the central rapidity region ($|\eta|<0.35$) and half in azimuth angle.
\cite{Adcox:2003zm}
PHENIX also has an excellent gamma triggering capability
(the threshold is 0.8~GeV or 1.4~GeV) which makes
high-statistics $\pi^0$ measurement feasible.\cite{Okada:1}
EMCal based trigger without coincidence with BBC is used
because the collision trigger efficiency based on BBC is low at $\sqrt{s}=62.4$ GeV.

The systematic uncertainty is evaluated
by the bunch shuffling technique,\cite{Adler:2004ps} and it is
found to be negligible.

\section{$A_{LL}$ calculation}
$\pi^0 A_{LL}$ $(A_{LL}^{\pi^0})$ is calculated by subtracting
$A_{LL}^{\textrm{\footnotesize BG}}$ from
$A_{LL}^{\pi^0 + \textrm{{\footnotesize BG}}}$.
$A_{LL}^{\pi^0 + \textrm{{\footnotesize BG}}}$ is the asymmetry
for the diphoton invariant-mass
range of 112~MeV/$c^2$-162~MeV/$c^2$ (under the $\pi^0$ peak).
$A_{LL}^{\textrm{\footnotesize BG}}$ is the asymmetry for the
range of 177~MeV/$c^2$-217~MeV/$c^2$ (higher side band).

\begin{figure}
  \includegraphics[height=.2\textheight]{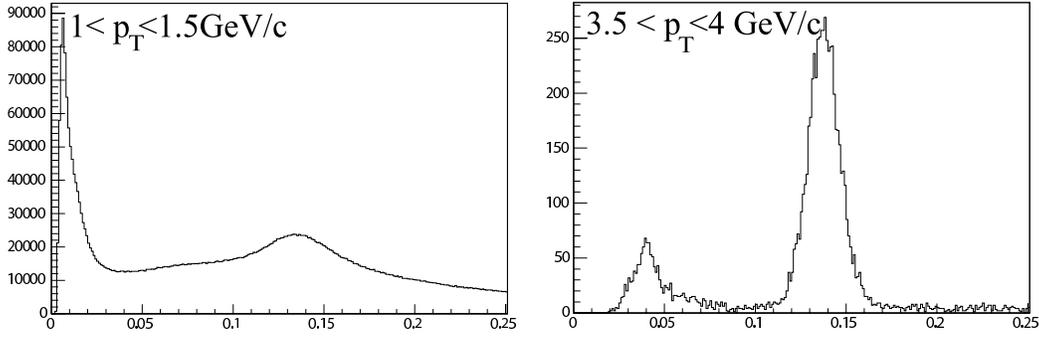}
  \caption{Diphoton invariant mass spectra.}
  \label{fig:inv}
\end{figure}

Figure \ref{fig:inv} shows the diphoton invariant mass spectra.
%
%
The lower mass peak corresponds to background
from hadrons and cosmic particles, which induce EMCal clusters with more
complicated structure, each of them are then splitted on several ones.
This peak roughly corresponds to two EMCal cell separation between two
clusters, which moves to higher mass with increasing cluster pair $p_T$.
Since we used EMCal based trigger without coincidence with collision trigger
at $\sqrt{s}=62.4$ GeV, the cosmic background is prominent
unlike in data at $\sqrt{s} = 200$ GeV.
The contribution of such background under $\pi^0$ peak is negligible
in the measured $p_T$ range. Since it does affect the lower side band,
the $A_{LL}^{\pi^0 + \textrm{{\footnotesize BG}}}$ estimation was done 
based only on the higher side band.
%
%
The subtraction is done by using the following formula.
\begin{equation}
A_{LL}^{\pi^0} = \frac{ A_{LL}^{\pi^0 + \textrm{{\footnotesize BG}}}
                  - rA_{LL}^{\textrm{\footnotesize BG}} }
	{1-r}
\end{equation}
where $r$ is the background fraction. 

\section{Results}
\begin{figure}
  \includegraphics[height=.3\textheight]{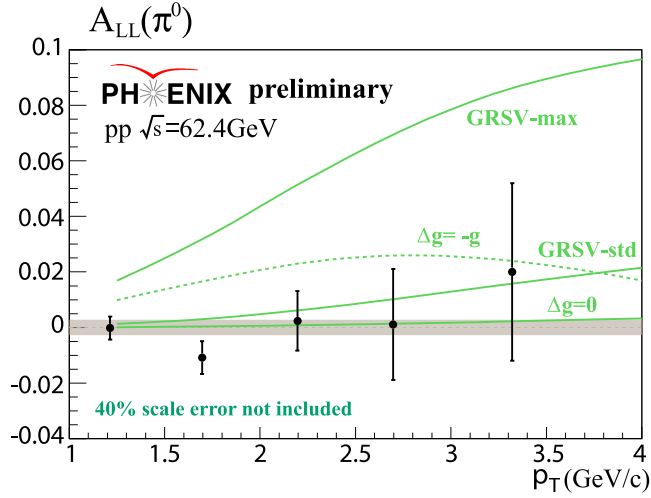}
  \caption{$\pi^0 A_{LL}$ as a function of $p_T$. The error bar denotes
  statistical uncertainty. Gray band denotes systematic error from
  relative luminosity.}
  \label{fig:allpi0_vs_pt}
\end{figure}

\begin{figure}
  \includegraphics[height=.3\textheight]{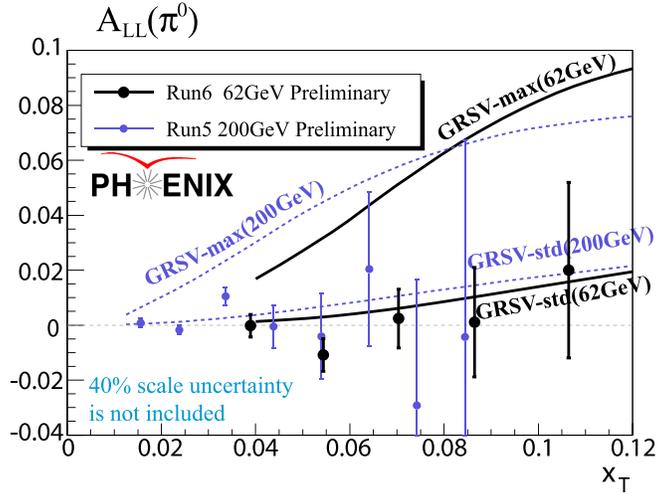}
  \caption{$\pi^0 A_{LL}$ as a function of $x_T$.}
  \label{fig:allpi0_vs_xt}
\end{figure}

Figure \ref{fig:allpi0_vs_pt} shows the Run 6 results of
$\pi^0 A_{LL}$ as a function of $p_T$.
$A_{LL}$ is consistent with zero over the measured $p_T$ region.
Detailed offline analysis on beam polarization is not provided yet
by the RHIC polarimeter group. Thus online values are used and
systematic uncertainty of 20\% is assigned for a single beam
polarization measurement. It introduces scaling uncertainty of
40\% on $A_{LL}$.
Theory curves based on pQCD using four proton spin models are also shown.\cite{Jager:2002xm}
The theory is based on pQCD; thus it is important to test pQCD applicability
at $\sqrt{s}=62.4$~GeV. To test pQCD applicability, analysis on 
$\pi^0$ cross-section is on-going.
With our cross section result,
we will be able to discuss our $A_{LL}$ result
further by comparing with pQCD calculations.
Figure \ref{fig:allpi0_vs_xt} shows the Run 6 results of
$\pi^0 A_{LL}$ as a function of
$x_T$ together with Run 5 results. A clear statistical improvement
can be seen in the large $x_T$ region.

\section{SUMMARY}
During the RHIC run in 2006, $\pi^0 A_{LL}$ at $\sqrt{s}=62.4$ GeV
was measured with the PHENIX detector. Preliminary results of 
$\pi^0 A_{LL}$ at $\sqrt{s}=62.4$~GeV with integrated luminosity of
60 nb$^{-1}$ and the average polarization of 48\% are presented.
There is a clear statistical improvement in the large $x_T$ regin compared
to the Run5 preliminary 
results at $\sqrt{s}=200$ GeV with integrated luminosity of 1.8pb$^{-1}$
and the average polarization of 47\%.
To extract the
gluon spin contribution to the proton, it is important
to test pQCD applicability at $\sqrt{s}=62.4$ GeV.
Analysis on cross-section is on-going to test pQCD at this energy.
With our cross section result,
we will be able to discuss our $A_{LL}$ result
further by comparing with pQCD calculations.





\bibliographystyle{aipproc}   

\bibliography{pi0spin2006}

\hyphenation{Post-Script Sprin-ger}
\begin{thebibliography}{7}
\expandafter\ifx\csname natexlab\endcsname\relax\def\natexlab#1{#1}\fi
\providecommand{\enquote}[1]{``#1''}
\expandafter\ifx\csname url\endcsname\relax
  \def\url#1{\texttt{#1}}\fi
\expandafter\ifx\csname urlprefix\endcsname\relax\def\urlprefix{URL }\fi
\providecommand{\eprint}[2][]{\url{#2}}

\bibitem[Boyle(2006)]{Boyle:2006ab}
K.~Boyle, \emph{AIP Conf. Proc.} \textbf{842}, 351--353 (2006),
  \eprint{nucl-ex/0606008}.

\bibitem[Hirai et~al.(2006)]{Hirai:2006sr}
M.~Hirai, S.~Kumano, and N.~Saito, \emph{Phys. Rev.} \textbf{D74}, 014015
  (2006), \eprint{hep-ph/0603213}.

\bibitem[Togawa et~al.(2007)]{Togawa:1}
M.~Togawa, et~al., \emph{RIKEN Accel. Prog. Rep. to be published} \textbf{40}
  (2007).

\bibitem[Adcox et~al.(2003)]{Adcox:2003zm}
K.~Adcox, et~al., \emph{Nucl. Instrum. Meth.} \textbf{A499}, 469--479 (2003).

\bibitem[Okada et~al.(2003)]{Okada:1}
K.~Okada, et~al., \emph{RIKEN Accel. Prog. Rep.} \textbf{36}, 248 (2003).

\bibitem[Adler et~al.(2004)]{Adler:2004ps}
S.~S. Adler, et~al., \emph{Phys. Rev. Lett.} \textbf{93}, 202002 (2004),
  \eprint{hep-ex/0404027}.

\bibitem[Jager et~al.(2003)]{Jager:2002xm}
B.~Jager, A.~Schafer, M.~Stratmann, and W.~Vogelsang, \emph{Phys. Rev.}
  \textbf{D67}, 054005 (2003), \eprint{hep-ph/0211007}.

\end{thebibliography}

\IfFileExists{\jobname.bbl}{}
 {\typeout{}
  \typeout{******************************************}
  \typeout{** Please run "bibtex \jobname" to optain}
  \typeout{** the bibliography and then re-run LaTeX}
  \typeout{** twice to fix the references!}
  \typeout{******************************************}
  \typeout{}
 }

\end{document}